\documentclass[]{spie}  

\usepackage{amsmath,amsfonts,amssymb}
\usepackage{graphicx}
\usepackage{hyperref}
\hypersetup{colorlinks,allcolors=blue}
\usepackage{makecell}
\usepackage{colortbl}
\usepackage{array}
\usepackage{afterpage}
\usepackage{multirow}

\usepackage[table]{xcolor}

\def\arcdeg{\mbox{$^\circ$}}%
\def\arcmin{\mbox{$^\prime$}}%
\def\arcsec{\mbox{$^{\prime\prime}$}}%

\definecolor{eric}{rgb}{0.7, 0.7, 1.0}

\title{The high-speed X-ray camera on AXIS: design and performance updates}

\author[a]{Eric D. Miller}
\author[a]{Catherine E. Grant}
\author[a]{Robert Goeke}
\author[a]{Marshall W. Bautz}
\author[b]{Christopher Leitz}
\author[b]{Kevan Donlon}
\author[c,d,e]{Steven W. Allen}
\author[c]{Sven Herrmann}
\author[f]{Abraham D. Falcone}
\author[a]{F. Elio Angile}
\author[c]{Tanmoy Chattopadhyay}
\author[b]{Michael Cooper}
\author[b]{Mallory A. Jensen}
\author[a]{Jill Juneau}
\author[a]{Beverly LaMarr}
\author[a]{Andrew Malonis}
\author[c,d]{R. Glenn Morris}
\author[c]{Peter Orel}
\author[c,e]{Abigail Y. Pan}
\author[g]{Steven Persyn}
\author[c]{Artem Poliszczuk}
\author[a]{Gregory Y. Prigozhin}
\author[b]{Ilya Prigozhin}
\author[h]{Andrew Ptak}
\author[i]{Christopher Reynolds}
\author[c,e]{Haley R. Stueber}
\author[b]{Keith Warner}
\author[c,e]{Daniel R. Wilkins}

\affil[a]{Kavli Institute for Astrophysics and Space Research, Massachusetts Institute of Technology, Cambridge, MA, USA}
\affil[b]{Lincoln Laboratory, Massachusetts Institute of Technology, Lexington, MA, USA}
\affil[c]{Kavli Institute for Particle Astrophysics and Cosmology, Stanford University, Stanford, CA, USA}
\affil[d]{SLAC National Accelerator Laboratory, Menlo Park, CA, USA}
\affil[e]{Department of Physics, Stanford University, Stanford, CA, USA}
\affil[f]{Department of Astronomy and Astrophysics, Pennsylvania State University, University Park, PA, USA}
\affil[g]{Southwest Research Institute, San Antonio, TX, USA}
\affil[h]{NASA Goddard Space Flight Center, Greenbelt, MD, USA}
\affil[i]{Department of Astronomy, University of Maryland, College Park, MD, USA}

\authorinfo{Further author information:\\Eric D.~Miller: E-mail: milleric@mit.edu}

\begin{document} 
\maketitle

\begin{abstract}
AXIS, a Probe mission concept now in a Phase A study, will provide transformative studies of high-energy astrophysical phenomena thanks to its high-resolution X-ray spectral imaging. These capabilities are enabled by improvements to the mirror design that greatly increase the X-ray throughput per unit mass compared to heritage missions like Chandra; and to the detector system, which operates more than an order of magnitude faster than heritage instruments while maintaining excellent spectral performance and low power consumption. We present updates to the design of the AXIS High-Speed Camera, a collaborative effort by MIT, Stanford University, the Pennsylvania State University, and the Southwest Research Institute. The camera employs large-format MIT Lincoln Laboratory CCDs that feature multiple high-speed, low-noise output amplifiers and an advanced single-layer polysilicon gate structure for fast, low-power clock transfers. A first lot of prototype CCID100 CCDs has completed fabrication and will soon begin X-ray performance testing. The CCDs are paired with high-speed, low-noise ASIC readout chips designed by Stanford to provide better performance than conventional discrete solutions at a fraction of the power consumption and footprint. Complementary Front-End Electronics employ state-of-the-art digital video waveform capture and advanced signal processing to further deliver low noise at high speed. The detector system is housed in a high-heritage camera body that ensures precise thermal control and minimization of molecular contamination. The Back-End Electronics provide high-speed identification of candidate X-ray events and transient monitoring that relays fast alerts of changing sources to the community. We highlight updates to our parallel X-ray performance test facilities at MIT and Stanford, and review the current performance of the CCD and ASIC technology from testing of prototype devices. These measurements achieve excellent spectral response at the required readout rate, demonstrating that we will meet mission requirements and enable AXIS to achieve world-class science.
\end{abstract}


\section{Introduction}
\label{sect:intro}

The Advanced X-ray Imaging Satellite (AXIS)\cite{Reynolds2023_AXIS} is a mission concept proposed in response to the NASA Astrophysics Probe Explorer call, and is now in Phase A competition for final selection for flight. AXIS provides high-throughput, high-spatial-resolution X-ray imaging spectroscopy, and it will allow ground-breaking studies addressing key science priority areas identified by the National Academies’ 2020 Decadal Survey on Astronomy and Astrophysics\cite{Astro2020}. In particular, AXIS will study the birth and evolution of super-massive black holes and the mechanisms of galactic feedback with a series of sensitive surveys; it will allow rapid follow-up of transient alerts to further our understanding of the dynamic Universe; and it will enable a rich General Observer program as a highly capable and flexible observatory. AXIS will provide the crucial X-ray counterpart to the panchromatic suite of large observatories of the 2030s, including JWST, Rubin, Roman, LIGO/Virgo/Kagra, LISA, SKA, and Euclid.

The high throughput and outstanding spatial resolution of AXIS present technical challenges for the detector system. To avoid photon pile-up, the camera must operate at frame rates significantly faster than similar detectors on Chandra and Suzaku, while retaining the excellent spectral performance of those heritage instruments across the full X-ray energy band. To achieve this performance, we take advantage of recent technical advances in the design and fabrication of the detectors and electronics chain. The AXIS camera design follows careful consideration of the baseline mission performance, shown in Table \ref{tab:axis_reqs} with the derived camera requirements.

\begin{table}[b]
\caption{AXIS baseline performance requirements relevant to the camera. \label{tab:axis_reqs}}
\footnotesize
\begin{center}       
\begin{tabular}{|l|l|} 
\hline\hline
\multicolumn{2}{|l|}{\textbf{AXIS mission parameters}} \\ \hline
Spatial resolution at 1 keV (HPD) & 1.5\arcsec\ (on-axis) \\
                              & 1.8\arcsec\ (FoV-average) \\\hline 
Effective area at 1 keV       & 3900 cm$^2$ (on-axis) \\
                              & 2990 cm$^2$ (FoV-average) \\\hline
Effective area at 6 keV       & 600 cm$^2$ (on-axis) \\
                              & 360 cm$^2$ (FoV-average) \\\hline
Field of view                 & 24\arcmin\ diameter \\\hline
Energy band                   & 0.3--10 keV \\\hline
Energy resolution (FWHM)      & $\leq$70 eV (at 0.5 keV) \\     
                              & $\leq$100 eV (at 1 keV) \\
                              & $\leq$150 eV (at 6 keV) \\\hline
Orbit                         & L2 halo orbit\\
Prime mission lifetime        & 5 years\\ \hline
\multicolumn{2}{|l|}{\textbf{High-Speed Camera characteristics}} \\ \hline
Frame rate                  & $\geq$ 5 fps (goal $\geq$ 20 fps)\\ \hline
Pixel size                  & 24 $\mu$m (0.55\arcsec) \\ \hline
Readout noise               & $\leq$ 3 e- RMS \\ \hline
Focal plane temperature     & $-110 \pm 0.1$\arcdeg C \\ \hline
\hline
\end{tabular}
\end{center}
\end{table} 

In this proceeding, we present updates to the design for the AXIS High-Speed Camera\cite{Miller2023_AXIS}, a collaborative effort of the MIT Kavli Institute for Astrophysics and Space Research (MKI), MIT Lincoln Laboratory (MIT/LL), Stanford University's Kavli Institute for Particle Astrophysics and Cosmology (KIPAC), the Pennsylvania State University, and the Southwest Research Institute (SwRI). Key updates include a detailed design for the AXIS CCD developed by MIT/LL, continuing a multi-year effort to develop fast, low-noise detectors for future strategic X-ray missions.\cite{Bautzetal2019,Prigozhinetal2020,Lamarretal2020, Bautzetal2020, LaMarretal2022, Prigozhinetal2022, Milleretal2022c, LaMarretal2022b, Bautzetal2022,Herrmannetal2020,Chattopadhyayetal2020,Oreletal2022,Herrmannetal2022,Chattopadhyay2022_ccd,Chattopadhyayetal2022,Chattopadhyay2023_sisero,LaMarr2024_SPIE,Bautz2024_SPIE,Herrmann2024_SPIE,Orel2024_SPIE,Chattopadhyay2024_SPIE}. We describe the X-ray performance test beds that have been developed and deployed in our parallel facilities at MIT and Stanford, and provide an updated summary of expected focal plane performance. We demonstrate that the advanced technology is on track to meet the baseline requirements of the AXIS mission.

\section{Camera design updates}
\label{sect:camera_design}

The updated High-Speed Camera retains most of the features of the previous design\cite{Miller2023_AXIS}. However, there has been a significant change to the mission profile: AXIS will now fly in an L2 halo orbit instead of the originally planned low-inclination, low-Earth orbit, primarily to simplify the thermal design of the observatory. The increased particle radiation environment has necessitated changes to the mechanical and thermal design of the camera, as described below. In addition, the initial detector design has been finalized, front-illuminated AXIS CCDs have been fabricated, and the Front-End Electronics design work is progressing.

\begin{figure}[t]
\begin{center}
\includegraphics[width=.75\linewidth]{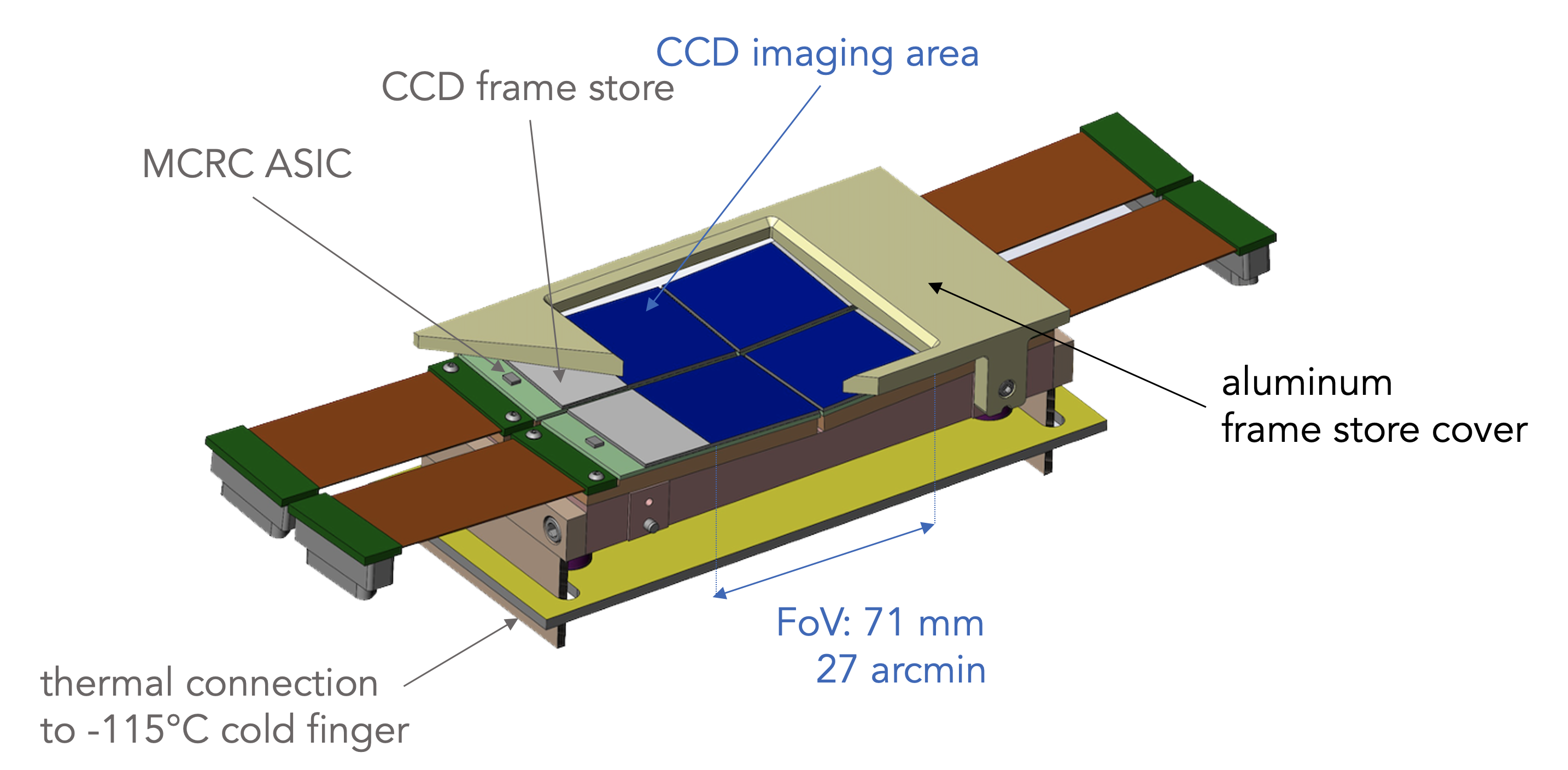}
\end{center}
\caption{AXIS detector array with components labeled. The array consists of four fast-readout frame-store CCDs arranged in a 2$\times$2 pattern, each with a dedicated ASIC readout chip co-mounted on the interposer. An aluminum cover shields the frame store regions from focused celestial X-rays during readout, and it provides the ASICs with additional particle damage protection. The entire array is maintained at -110\arcdeg C using a cold finger connection and active trim heaters.}
\label{fig:detarray}
\end{figure} 

\subsection{Camera signal chain}
\label{sect:signal_chain}

\subsubsection{Detector development}
\label{sect:detectors}

\begin{figure}[p]
\begin{center}
\includegraphics[width=.60\linewidth]{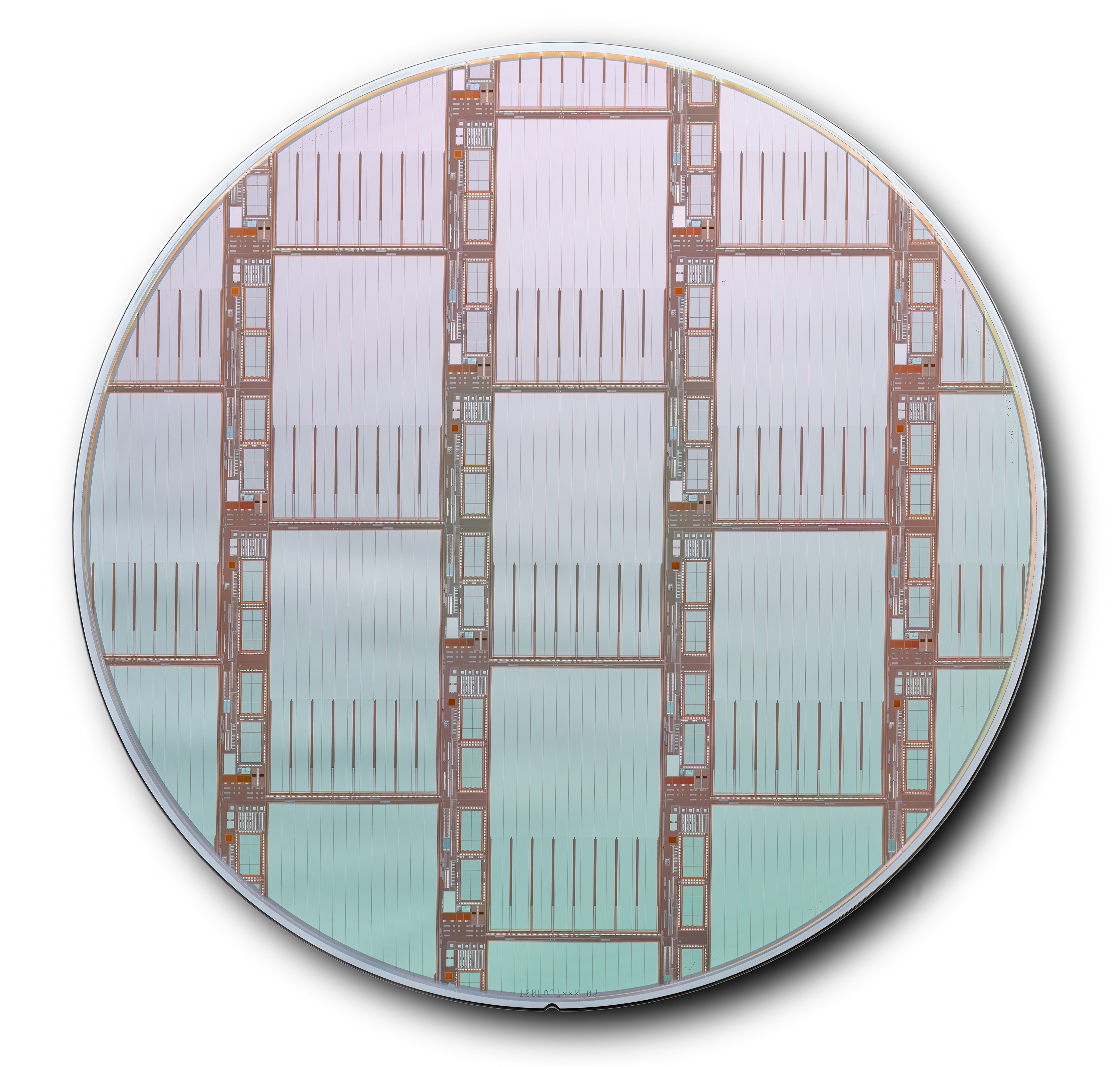}
\includegraphics[width=.35\linewidth]{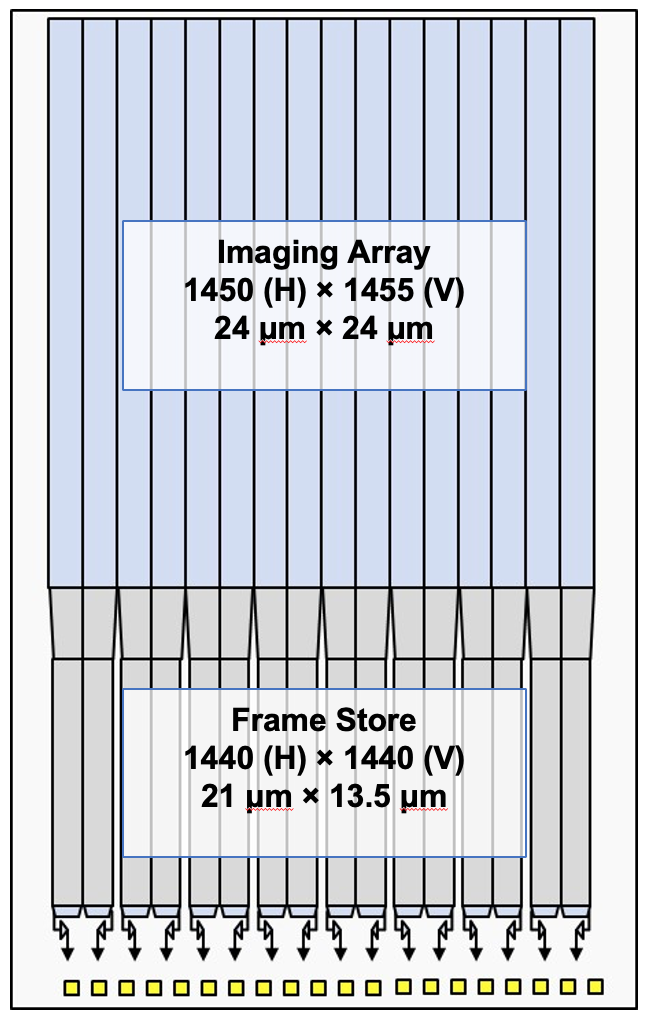}
\end{center}
\caption{(left) Photograph of a 200-mm wafer containing MIT/LL CCID100 CCDs through front-illumination. Each wafer is large enough to fabricate seven complete CCID100 CCDs. (right) Schematic showing the layout of the CCID100 CCD.}
\label{fig:ccid100_wafer}
\end{figure}

The AXIS focal plane (see Figure \ref{fig:detarray}) has four detectors arranged in a 2x2 array, spanning a region 27\arcmin$\times$27\arcmin\ on the sky. Each detector is a back-illuminated, frame-store CCD designated the CCID100 by the Advanced Imager Technology Group\footnote{\url{https://www.ll.mit.edu/r-d/advanced-technology/advanced-imager-technology}} at MIT/LL, who have designed them specifically for AXIS. A first lot of front-illuminated CCID100 devices has recently completed fabrication at MIT/LL\cite{Leitz2025_SPIE}; a photograph of a completed wafer is shown in Figure \ref{fig:ccid100_wafer}, along with a schematic of the CCID100. The imaging area has 1440$\times$1440 active pixels, each 24 $\mu$m and spanning 0.55\arcsec, sufficient to sample the AXIS PSF in all locations in the field of view. An advanced single-layer polysilicon gate structure allows for fast, low-power clock transfers at parallel rates of 1 MHz. Sixteen high-speed, low-noise, p-channel junction field-effect transistor (pJFET) outputs operating at 2 MHz on each CCD allow frame rates of $\sim$14 fps, running significantly faster than our baseline requirement of 5 fps and reducing pile-up of X-rays in individual frames. The CCDs are thinned to 100 µm and treated with a molecular beam epitaxy (MBE) process to passivate the entrance window and provide excellent spectral performance across the entire AXIS energy band. A thin layer of aluminum is directly deposited on each CCD to block optical and UV photons. To minimize dark current and ensure the best performance in the radiation environment at L2, the focal plane will be operated at $-$110$^{\circ}$C. 

\begin{figure}[p]
\begin{center}
\includegraphics[width=.40\linewidth]{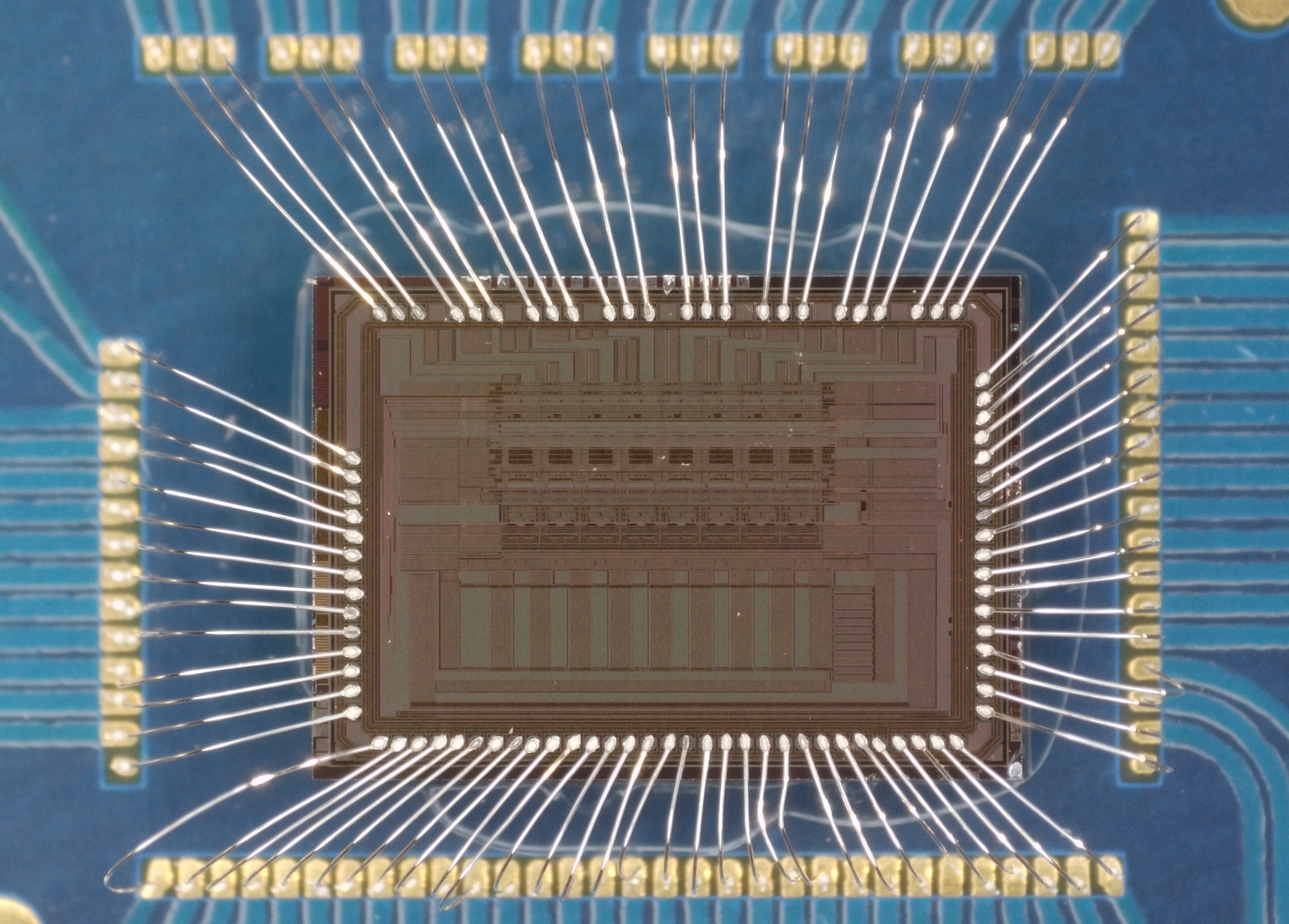}
\hspace*{.2in}
\includegraphics[width=.40\linewidth]{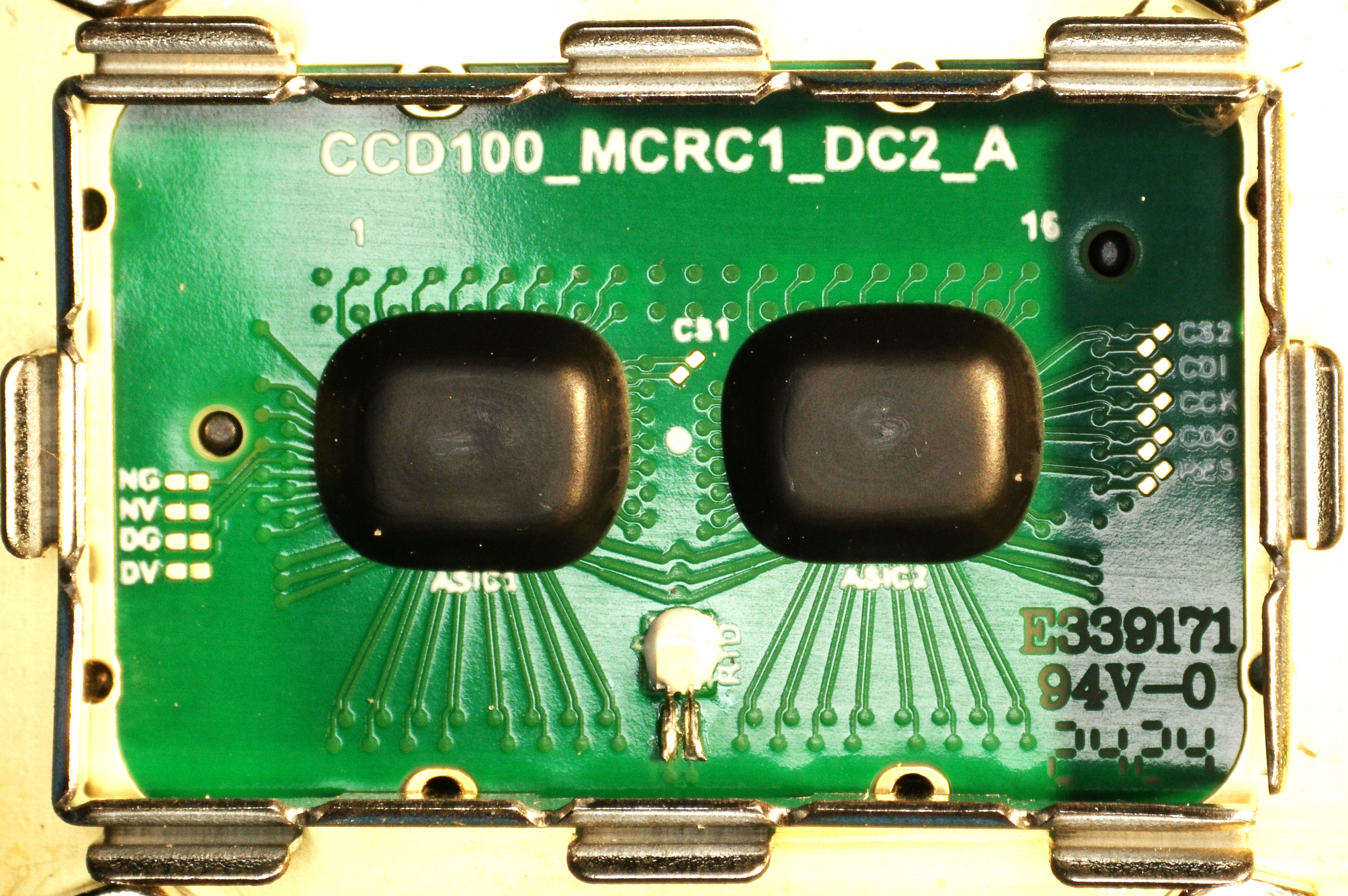}
\end{center}
\caption{(left) Photograph of a Multi-Channel Readout Chip (MCRC) ASIC with 8 channels. The chip itself is just 4.2 mm $\times$ 2.9 mm in size. (right) Dual 8-channel MCRC-V1 ASICs are mounted on a circuit board to support operation with the 16-channel CCID100 detector, allowing readout rates of up to 20 fps.}
\label{fig:mcrc}
\end{figure} 

Each CCD is integrated with a Multi-Channel Readout Chip (MCRC) application-specific integrated circuit (ASIC) to provide high-speed pre-amplification and buffering at a fraction of the power and mechanical footprint of conventional discrete electronics. The MCRC, developed by Stanford University and shown in Figure \ref{fig:mcrc}, features a capacitively coupled voltage amplifier and provides a fully differential output signal to the downstream electronics. The design and performance of the 8-channel AXIS prototype MCRC-V1 are described in detail elsewhere\cite{Orel2024_SPIE}. Each CCD+ASIC assembly is mechanically and electrically independent from the others, so a loss of one detector has no impact on the remainder of the focal plane.

CCDs from the completed front-illuminated CCID100 wafer lot will soon undergo testing and X-ray performance characterization with MCRC-V1 ASICs at our parallel facilities in the X-ray Detector Lab\footnote{\url{https://sites.mit.edu/xraydetectorlab}} at MIT and in the  X-ray Astronomy and Observational Cosmology (XOC) group\footnote{\url{https://xoc.stanford.edu}} at Stanford, as described in Section \ref{sect:test_beds}. Results from other prototype CCDs using the MCRC-V1 ASIC are summarized in Section \ref{sect:ccdperf}.

\subsubsection{Front-End Electronics development}
\label{sect:fee}

\begin{figure}[t]
\begin{center}
\includegraphics[width=.80\linewidth]{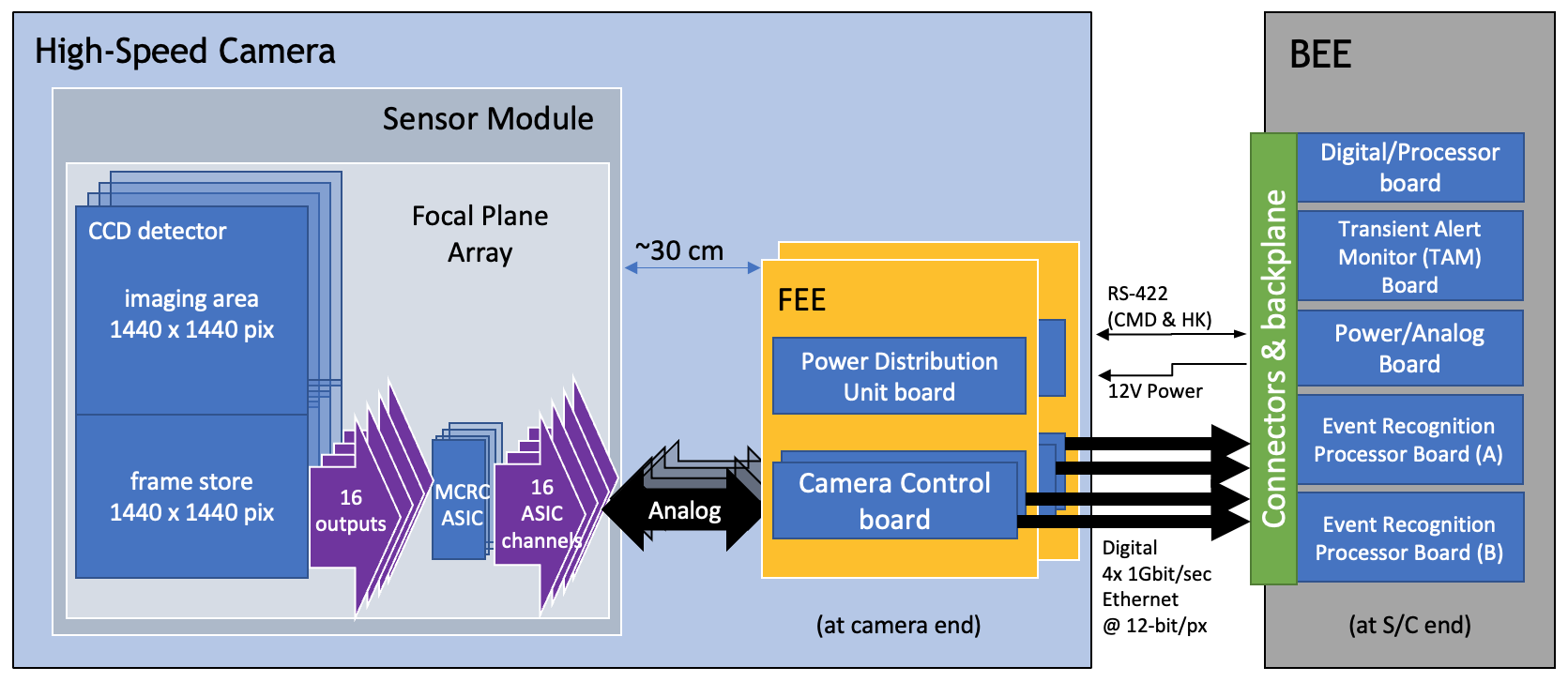}
\end{center}
\caption{Simplified block diagram of the camera system, showing the basic signal chain between the Focal Plane Array housed in the Sensor Module, the Front-End Electronics (FEE), and the Back-End Electronics (BEE).}
\label{fig:block_diagram}
\end{figure}

The Front-End Electronics (FEE), developed jointly by MIT and Stanford, provides power and clock voltages to the focal plane, amplifies and digitizes the output video signal, and controls all mechanisms and heaters on the camera. Two major changes have been incorporated into the FEE since our original design\cite{Miller2023_AXIS}. First, each CCD+ASIC is now controlled and read by a single Camera Control (CC) board. This design refinement allows for greater resiliency with no additional complexity: a loss of a single CC board will not affect the other three detector chains, and the boards are identical, allowing for efficient use of components and spares. Each CC board employs a Microchip PolarFire FPGA and Microchip ADCs to perform state-of-the-art digital video waveform capture at 50 Msamples/s, delivering low noise at high speed. The digitized video waveform is processed in the CC board to generate a raw image. Each board then sends the raw image data along an independent high-speed Ethernet link to the Back-End Electronics (BEE), which performs X-ray event finding and reduces the telemetry stream by several orders of magnitude. The BEE has been designed by Penn State and SwRI, and it is similar to what has been presented previously.\cite{Miller2023_AXIS}. A simplified block diagram of the camera system is shown in Figure \ref{fig:block_diagram}.

The second major change splits the FEE into two boxes. This is required to accommodate the short maximum run ($\leq$ 50 cm) of the flex cables carrying the analog signal from the focal plane to the FEE. Each box incorporates two CC boards to operate half of the focal plane, and each contains an identical Power Distribution Unit (PDU) board that provides power to the CC boards, operates the heaters, and controls camera mechanisms. Either board can accomplish these tasks, so a single FEE box could in principle fail without losing the entire focal plane. Each box and the boards it contains are identical, further reducing complexity.

\clearpage

\subsection{Mechanical design}

\begin{figure}[t]
\begin{center}
\includegraphics[width=.80\linewidth]{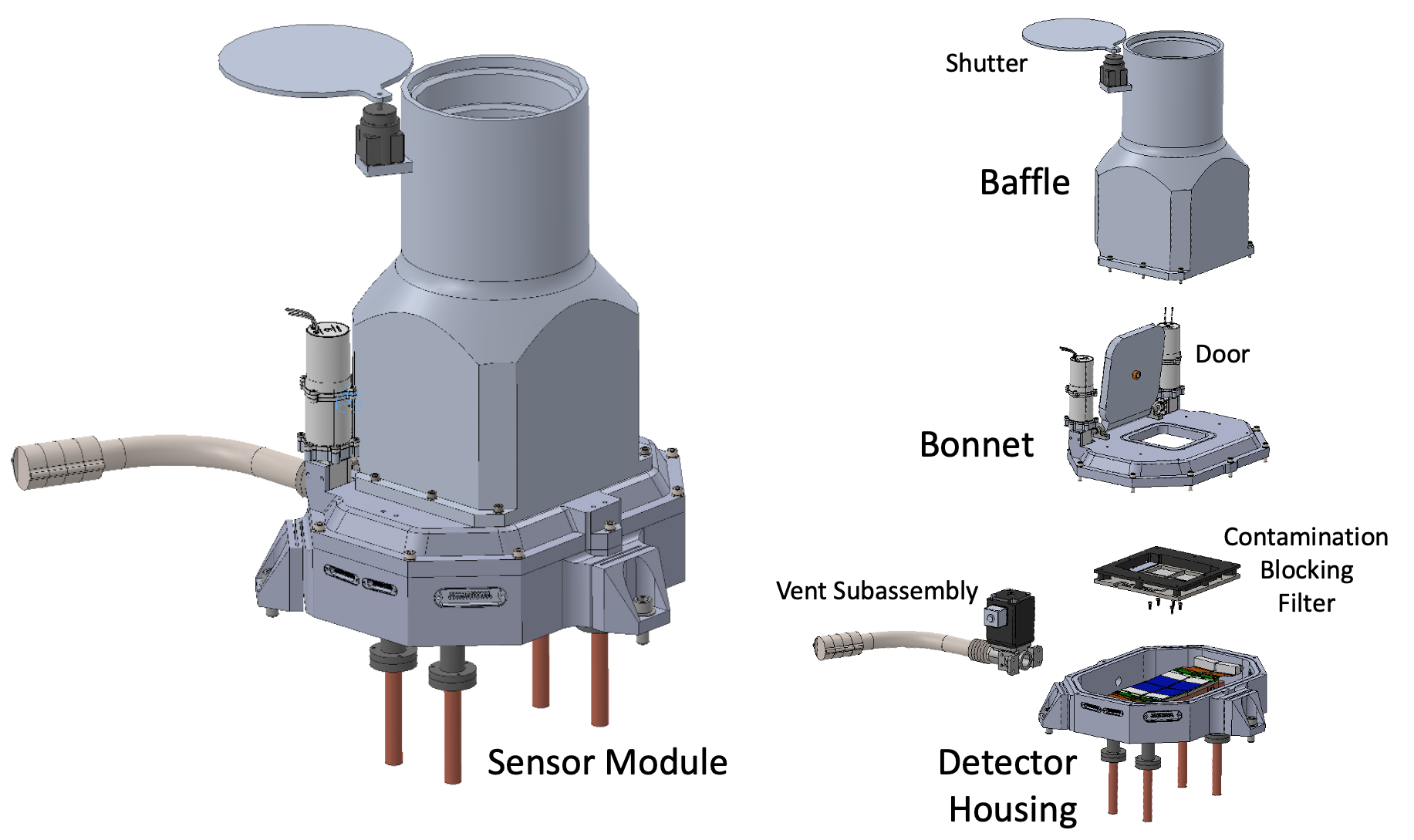}
\end{center}
\caption{The AXIS focal plane camera assembly, rendered as integrated (left) and in an exploded view (right). Various features are labeled and described further in the text.}
\label{fig:sensor_module}
\end{figure}

The focal plane array is housed in the Sensor Module, which provides a stable thermal environment and incorporates several mechanisms and other components required for optimal camera performance. The design of the camera, shown in Figure \ref{fig:sensor_module}, is similar to that presented previously\cite{Miller2023_AXIS}, with two major changes. First, the camera door has been redesigned to allow for automated open and close using a stepper motor. This change from a spring-loaded, single-open door simplifies integration and testing, as human intervention is not required to close the door and additionally we can close it under vacuum. The door houses an $^{55}$Fe calibration source that, when closed, illuminates the entire focal plane with 5.9 keV photons. This enables aliveness tests and performance monitoring during ground integration and on-orbit commissioning. The source is shielded to prevent focal plane illumination once the door opens.

The vacuum door, along with the vent subassembly, ensures that we can keep the volume within the detector housing under vacuum at all times after assembly. This protects the fragile contamination blocking filter (CBF) from acoustic loads, and it reduces the risk of molecular contamination building up on the cold focal plane surface. This volume is vented to space after launch, and once the door opens (expected to be a one-time operation), the CBF is maintained at a temperature above 0°C to provide additional protection against hydrocarbon contaminants condensing along the optical path. The CBF is composed of thin layers of aluminum and polyimide mounted on a 95\% open stainless-steel mesh, providing additional optical and UV light blocking to the aluminum layer on the CCDs.

The second major change is the addition of a commandable shutter at the top of the baffle. This shutter, designed as a 4-mm-thick aluminum blade rotated into the light path using a solenoid actuator, has been added as a result of the move to L2 and the associated increase in the particle background compared to LEO. The primary purpose of the shutter is to protect the focal plane from solar energetic particles during periods of high solar activity. The secondary purpose is to allow pristine observations of the non-X-ray background (NXB) without focused celestial X-rays, similar to XMM-Newton EPIC and eROSITA filter-wheel-closed observations or the ``stowed'' background observations by Chandra ACIS. The constraints on the spectral, spatial, and temporal variability afforded by these measurements will enhance the science return for faint, diffuse emission sources. The strategy to employ the shutter for NXB measurements is currently under study by the AXIS science team.

Several minor changes to the Sensor Module have resulted from the different thermal environment at L2 and from the detailed thermal design work accomplished during Phase A. These include a modified CBF frame design and a modified thermal interface between the focal plane and the radiator cold straps. The latter will provide a stable $-$115°C connection, and we will use trim heaters to maintain a focal plane temperature of $-$110$\pm$0.1°C. The redesigned L2 camera cooling system is passive, another change from the thermo-electric-controlled cooling in LEO.

\section{Camera performance updates}

\subsection{Summary of AXIS prototype CCD designs}

\begin{figure}[p]
\begin{center}
\includegraphics[height=2in]{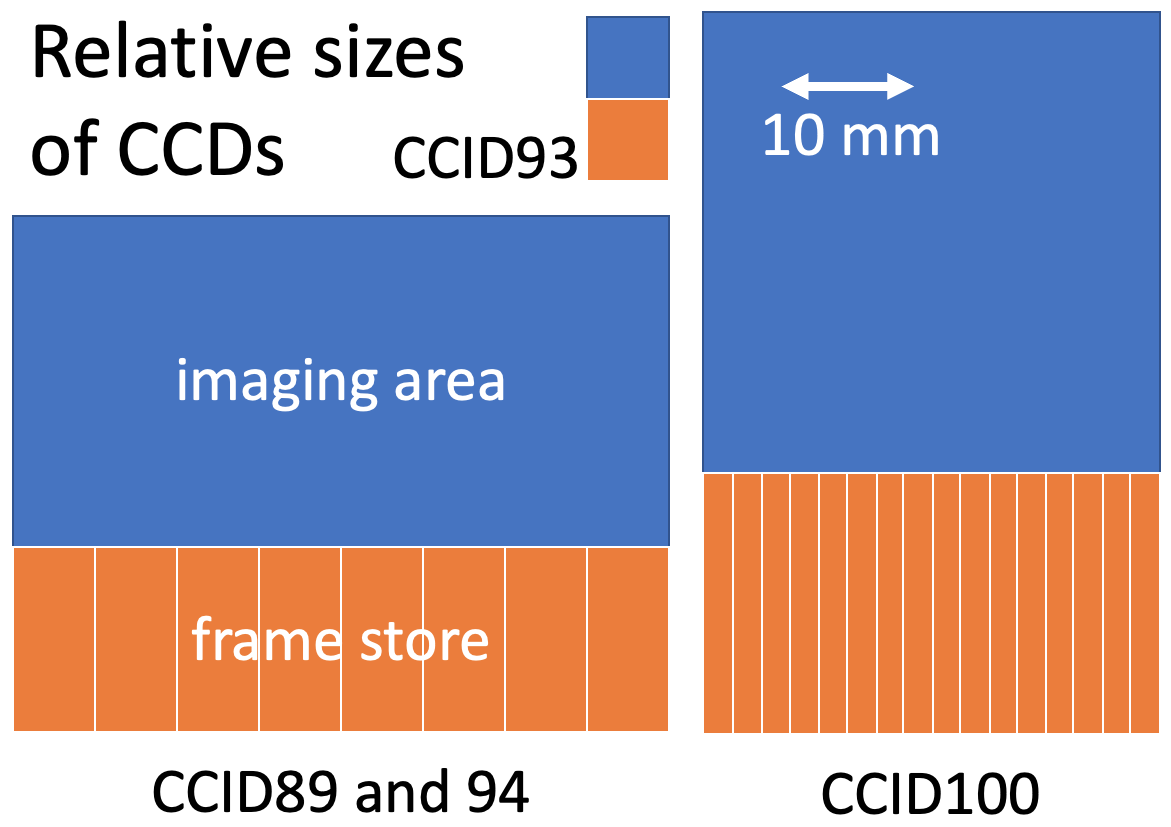}
\hspace*{.2in}
\includegraphics[height=2.5in]{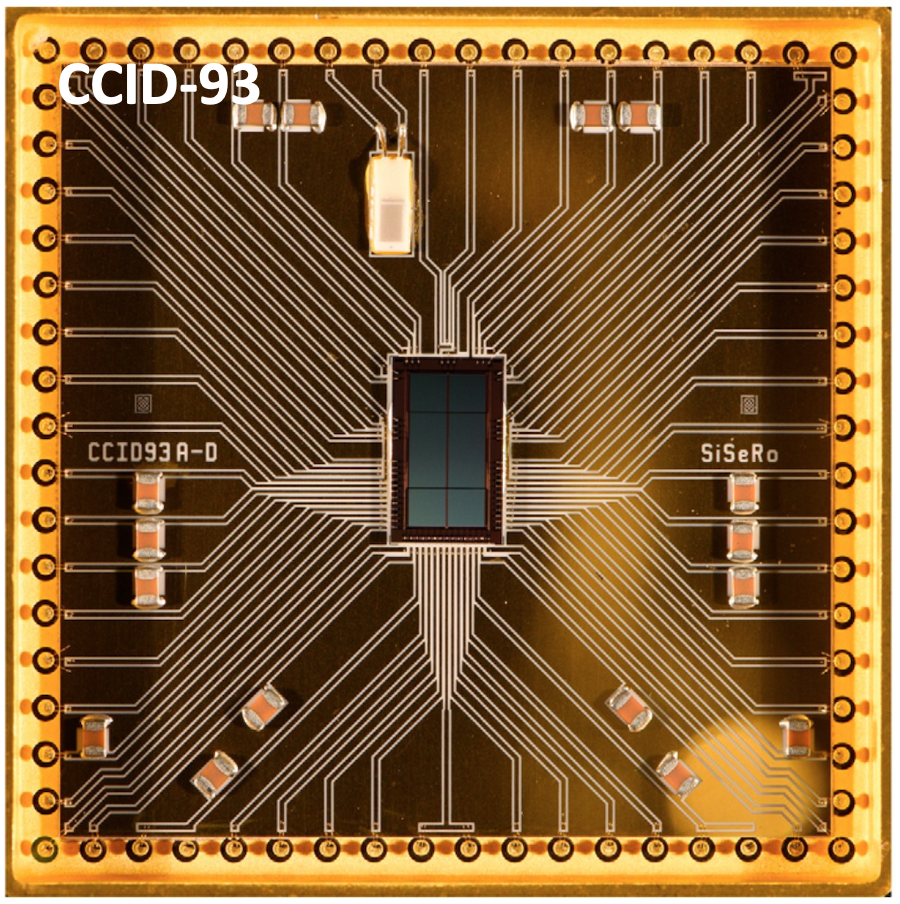}\\
\vspace*{.2in}
\includegraphics[height=3in]{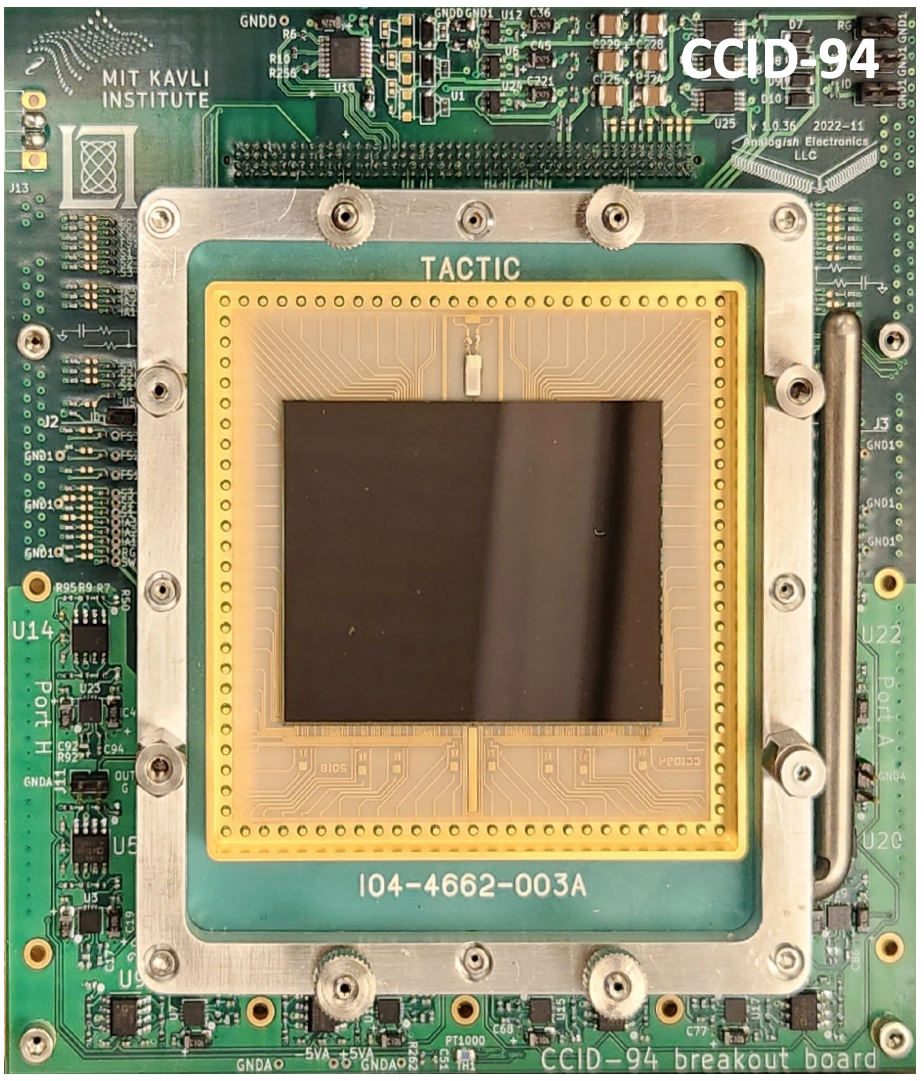}
\hspace*{.2in}
\includegraphics[height=3in]{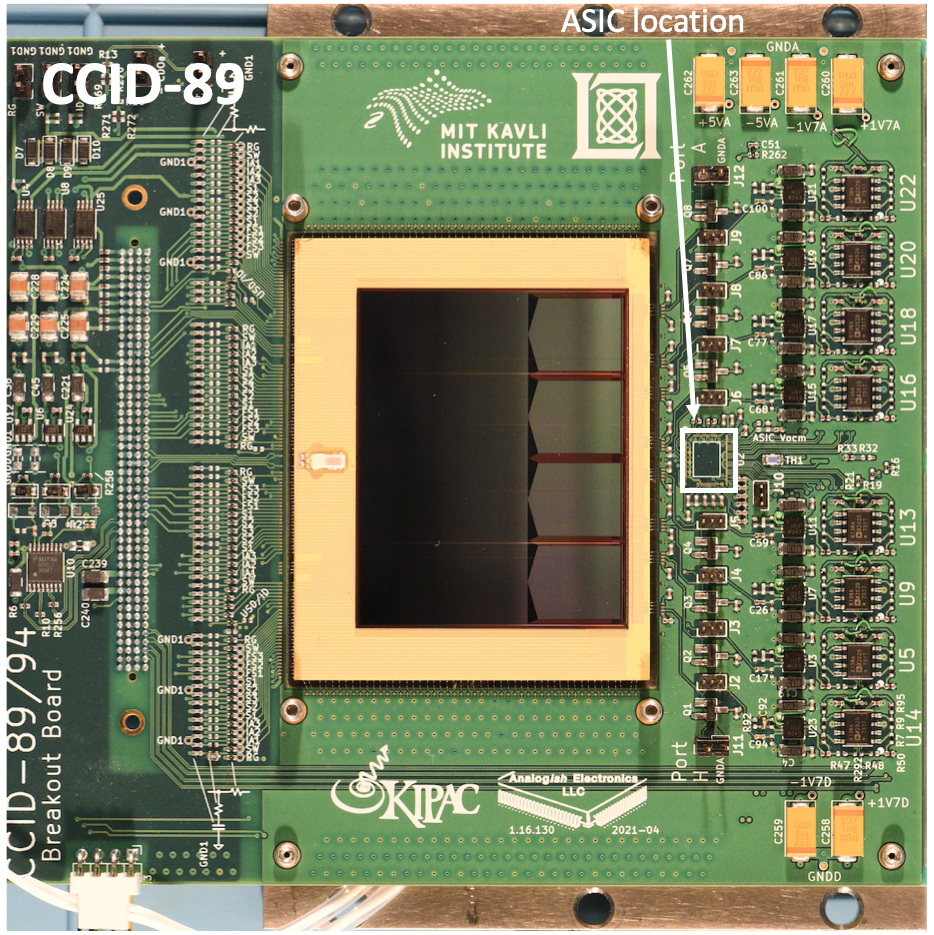}
\end{center}
\caption{Photographs of packaged MIT/LL CCDs under testing for AXIS development and performance demonstration. Photos are not at the same scale; the schematic in the upper left panel shows the relative sizes, and the number of output nodes are shown as orange segments in the frame store region. The CCID94 and 89 are the same size but shown in different orientations. The CCID89 shown is a front-illuminated device, so the metal layer delineating the frame store regions is visible, while it cannot be seen in the back-illuminated CCID94. The CCID100 is the AXIS design CCD and has not yet been delivered to our X-ray test facilities; a CCID100 wafer photo is shown in Figure \ref{fig:ccid100_wafer}.}
\label{fig:protoccds}
\end{figure}

Our team has been active in recent years characterizing the X-ray performance of several versions of MIT/LL CCDs that can be considered prototypes for AXIS. These detectors have been described previously\cite{Miller2023_AXIS}; however, we also summarize them here in Table \ref{tab:protoccds} as there are some slight changes to the CCID100 design. Figure \ref{fig:protoccds} shows the relative size of each CCD along with photographs of the detectors in their test packages, except for the CCID100, which has not yet been delivered to our test facilities. While the CCID93 and CCID94 are useful devices for characterizing certain features of the AXIS CCID100, the CCID89 is identical in most ways to the CCID100 design, with the same pixel size, same number of pixels, single-poly gate structure, and pJFET readouts. We focus on the X-ray performance of this device in the sections below. Test results demonstrating key performance metrics using other prototype devices operating with the MCRC-V1 ASIC are presented elsewhere\cite{Orel2024_SPIE}, including a report on tuning the CCID93 and ASIC in our Stanford test bed to achieve optimal spectral response at 6 keV.\cite{Stueber2025_SPIE}

\begin{table}[t]
\caption{Features of MIT/LL CCDs under testing for AXIS at MIT and Stanford, compared to the AXIS CCD design.}\label{tab:protoccds}
\begin{center}       
\scriptsize
\begin{tabular}{|p{1in}|>{\centering}p{1.2in}|>{\centering}p{1.14in}|>{\centering}p{1.14in}|>{\centering\arraybackslash}p{1.35in}|}
\hline\hline
\textbf{Feature}                   & \textbf{CCID93} & \textbf{CCID94} & \textbf{CCID89} & \textbf{CCID100 (AXIS)} \\ \hline
Format                    & Frame-transfer, 512$\times$512 pixel imaging array                       & \multicolumn{2}{|c|}{Frame-transfer, 2048$\times$1024 pixel imaging array} & Frame-transfer, 1440$\times$1440 pixel imaging array \\ \hline
Image area pixel size  & 8$\times$8 $\mu$m                                                            & 24$\times$24 $\mu$m                & 24$\times$24 $\mu$m             & 24$\times$24 $\mu$m \\ \hline
Output ports              & 1 pJFET, 1 SiSeRO (independent)                                     & 8 pJFET                   & 8 pJFET                & 16 pJFET \\ \hline
Transfer gate design      & Single layer polysilicon                                            & Triple layer polysilicon  & Single layer polysilicon & Single layer polysilicon \\ \hline
Additional features       & Regions with 0.5, 1, 2 $\mu$m and no trough; charge injection & Trough, charge injection          & Trough, charge injection       & Trough, charge injection\\ \hline
BI detector thickness     & 50 $\mu$m                                                               & 50 $\mu$m                     & 50 $\mu$m                  & 100 $\mu$m \\ \hline
Back surface              & MIT/LL MBE 5--10 nm      & MIT/LL MBE 10 nm          & MIT/LL MBE 5--10 nm    & MIT/LL MBE 5--10 nm \\ \hline
Typical serial rate       & 2--5 MHz                                                           & 0.5 MHz                   & 2--5 MHz              & $\geq$2 MHz \\ \hline
Typical parallel rate     & 0.1 MHz                                                             & 0.1 MHz                   & 0.2 MHz                & $\geq$0.5 MHz\\ \hline
Full frame read time   & 0.15--0.56 s                                                       & 1 s                       & 0.15--0.56 s          & $\leq$0.2 s \\
\hline\hline
\end{tabular}
\end{center}
\end{table} 

\subsection{Updates to X-ray performance test facilities}
\label{sect:test_beds}

\begin{figure}[t]
\begin{center}
\includegraphics[width=6in]{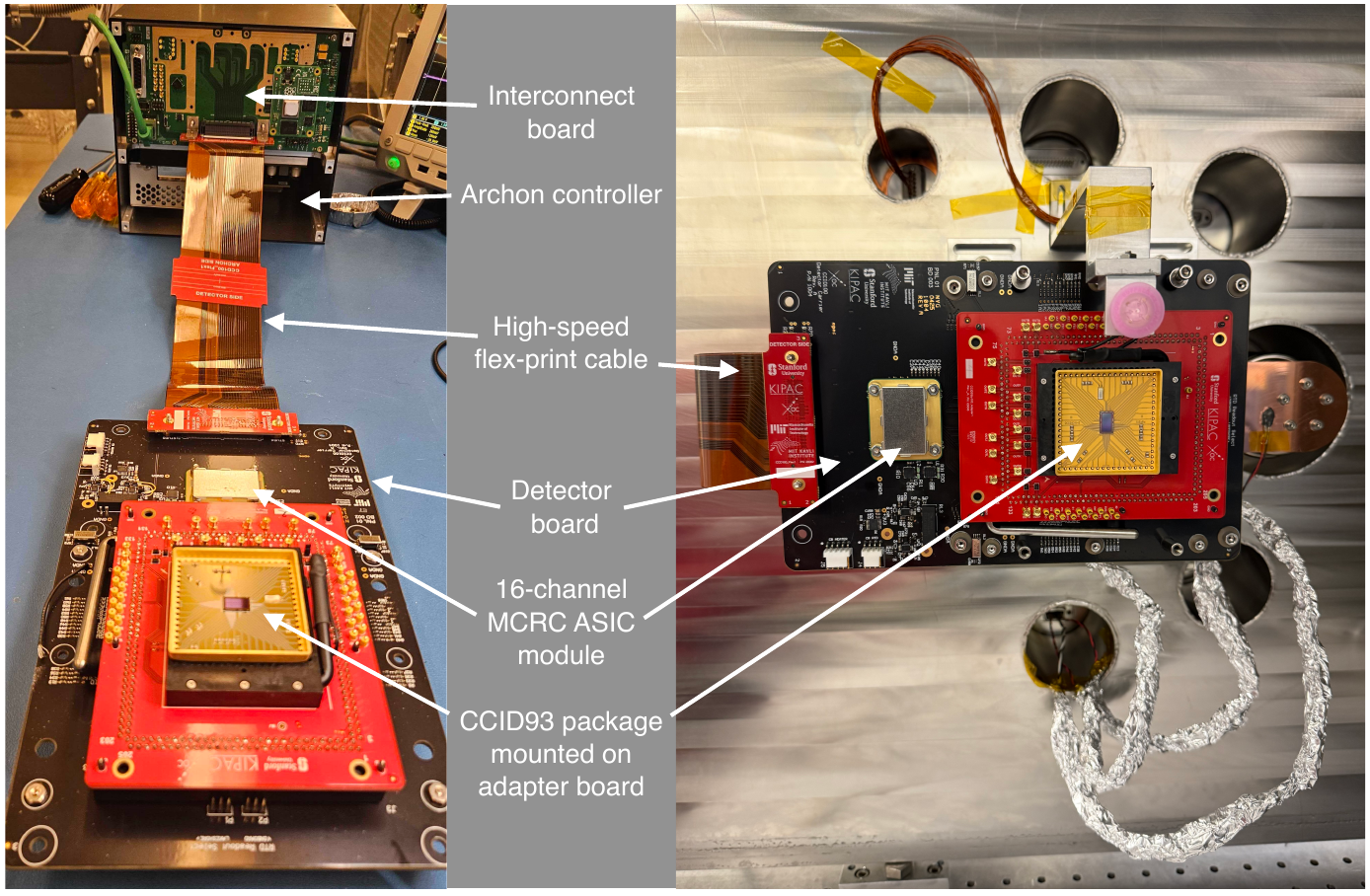}
\end{center}
\caption{(left) CCID100 test bed circuit boards arranged on the bench top in the X-ray Detector Lab at MIT. This setup includes a 104-pin CCID93 package inserted in a ZIF socket on the red adapter board, all mounted on the black CCID100 detector board. This allows testing of the board and ASIC functions before the CCID100 is installed. The 16-channel MCRC ASIC module is installed on the detector board. The flex print leads to the interconnect board, which connects directly to the Archon controller. (right) An identical set of boards installed in the vacuum housing in the XOC lab at Stanford. The interconnect board and Archon are outside of the vacuum chamber and therefore not shown. The round pink cover shows the location of a silicon drift detector (SDD) used for beamline monitoring.}
\label{fig:testbeds}
\end{figure} 


To incorporate the 16-channel CCID100 and dual-packaged 8-channel MCRC-V1 ASIC module for X-ray performance testing, the groups at MIT and Stanford have designed a common signal-chain circuitry that can be deployed in both facilities, allowing for standardized testing of multiple devices. The design is described elsewhere in these proceedings\cite{Pan2025_SPIE} and is shown in Figure \ref{fig:testbeds}. In brief, it incorporates a zero-insertion force (ZIF) socket for straightforward swapping of CCID100s mounted in 261-pin Kovar packages; a socket for simplified installation of a 16-channel MCRC ASIC module; and a flex-print connection between the in-vacuum detector board and the interconnect board that interfaces with the bench-top Archon CCD controller\footnote{\url{http://www.sta-inc.net/archon}}, the latter providing power and clock signals. The ASICs' digital core registers are programmed through a serial peripheral interface (SPI) bus using a bench-top computer. 

Recent tests using an adapter board to accommodate the 104-pin CCID93 package have allowed us to verify that the board, including the dual-ASIC module, functions as designed in both our test facilities. We have also successfully commissioned the mechanical, vacuum, and cooling systems in the Stanford test bed using the CCID100 board set-up.

Work is ongoing to develop a prototype FEE CC board to incorporate in our CCID100 testing. In particular, we will demonstrate performance of the high-speed waveform digitization required for AXIS using commercial versions of our selected flight FPGA and ADC components (see Section \ref{sect:fee}). We are currently designing modifications to the test bed circuitry to allow this prototype board to intercept the output video signals. This mode will use the Archon to provide power and clocking signals to the CCDs and ASICs. We anticipate incorporating this prototype FEE into our CCID100 X-ray performance testing within the next year.

\subsection{X-ray performance measurements}
\label{sect:ccdperf}

X-ray testing of prototype CCDs across the AXIS 0.3--10 keV band has been extensively reported in previous work\cite{Bautzetal2019,Prigozhinetal2020,Lamarretal2020, Bautzetal2020, LaMarretal2022, Prigozhinetal2022, Milleretal2022c, LaMarretal2022b, Bautzetal2022,Herrmannetal2020,Chattopadhyayetal2020,Oreletal2022,Herrmannetal2022,Chattopadhyay2022_ccd,Chattopadhyayetal2022,Chattopadhyay2023_sisero,LaMarr2024_SPIE,Bautz2024_SPIE,Herrmann2024_SPIE,Orel2024_SPIE,Chattopadhyay2024_SPIE}. We here summarize the noise and spectral response of a back-illuminated CCID89 detector studied in the X-ray Detector Lab at MIT. These performance metrics are directly tied to the baseline requirements shown in Table \ref{tab:axis_reqs}, and the CCID89 is nearest in design to the AXIS CCID100, so these measurements are most relevant for demonstrating that we meet AXIS requirements.

Noise introduced by the readout chain hampers our ability to accurately reconstruct the total energy of detected photons, since each pixel in a multi-pixel event must be summed to recover the photon energy and each pixel contributes noise. The effect is worst in the soft band, since we must impose a noise threshold during this summation, and any real signal lost below threshold degrades our energy resolution. In the worst case, entire events may be lost below threshold, leading to a reduction in QE.

\begin{figure}[p]
\begin{center}
\includegraphics[width=4.5in]{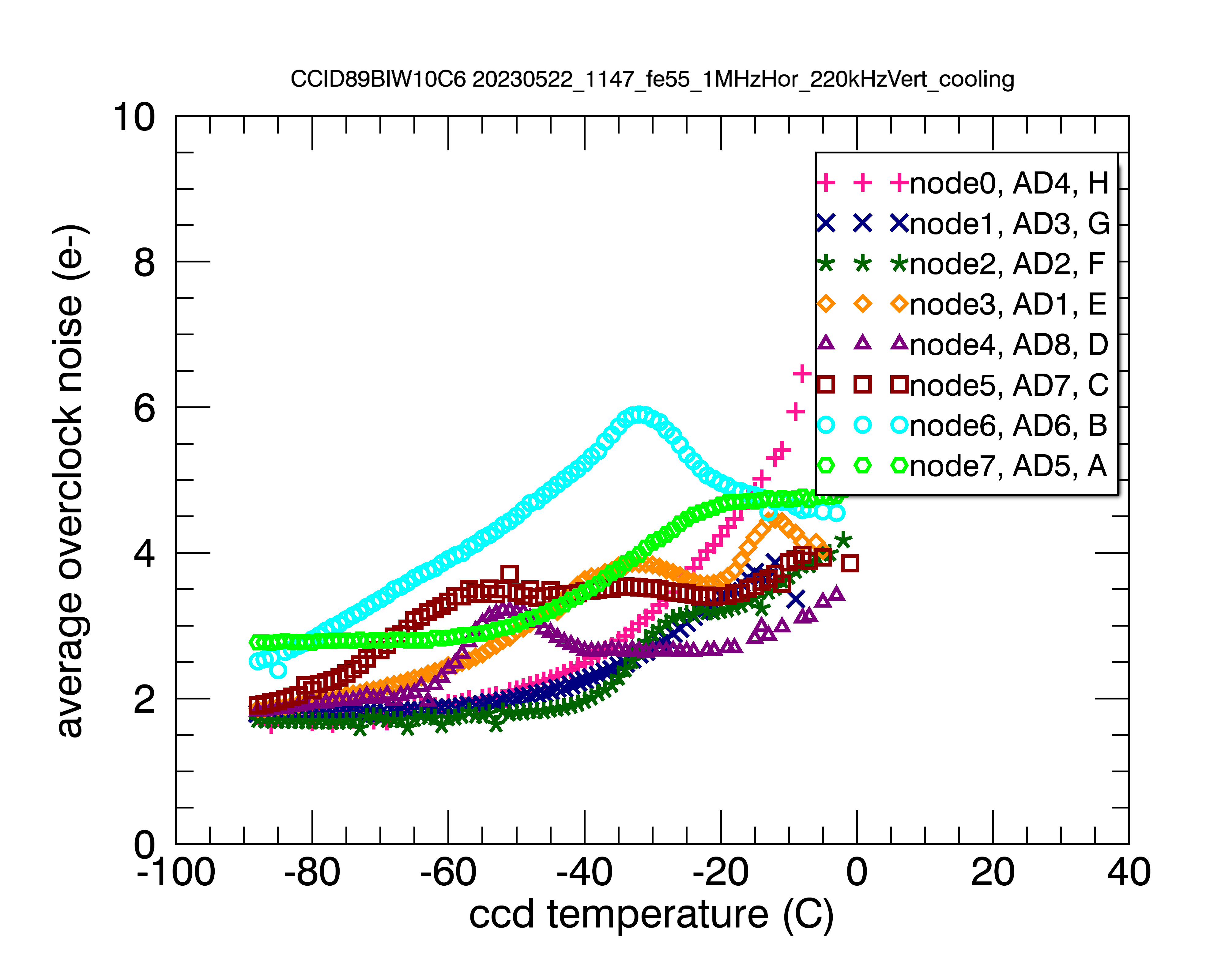}
\end{center}
\caption{Noise as a function of temperature for a back-illuminated CCID89. Each of the eight readout nodes is shown in a different color, and they all meet the AXIS requirement of $\leq$ 3 e- RMS at the lowest temperature achieved in this test.}
\label{fig:ccid89_noise}
\end{figure}

Our noise measurements indicate that these prototype devices meet the requirement of $\leq 3$ e- RMS, as shown in Figure \ref{fig:ccid89_noise}. While these measurements do not extend to the $-$110°C operating temperature of the AXIS focal plane at L2, testing of other devices with similar pJFET outputs indicates similar behavior at lower temperatures. Importantly, all eight output nodes on this device meet the noise requirement while operating at 1 MHz, sufficient to achieve 7 fps on the 16-channel CCID100.

\begin{figure}[p]
\begin{center}
\includegraphics[width=4.5in]{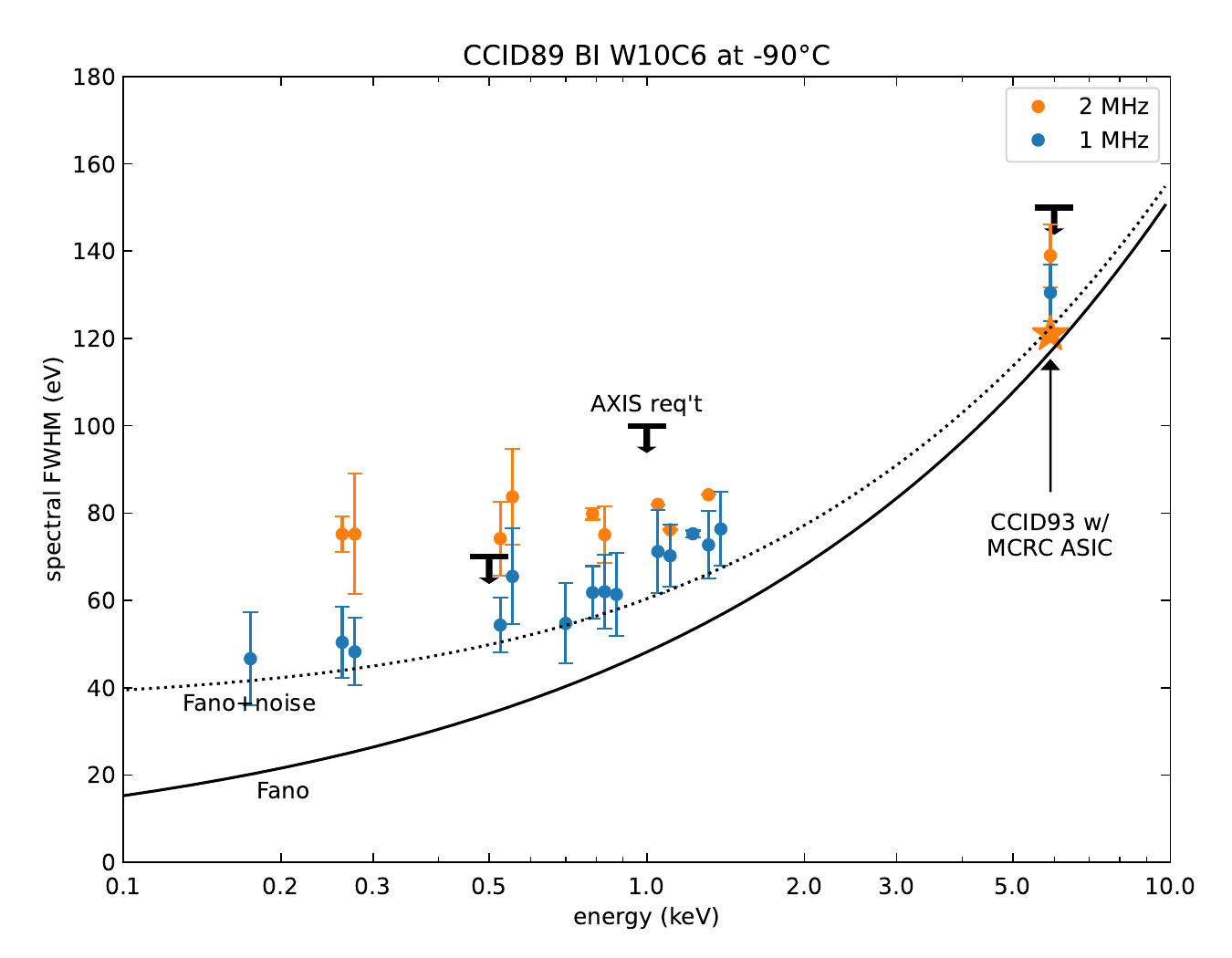}
\end{center}
\caption{Spectral resolution as a function of energy for an AXIS prototype back-illuminated CCID89 CCD running at two different output speeds. The data were obtained on the In-Focus Monochromator (IFM) in our facility at MIT, except for the measurement at 5.9 keV taken with a CCID93 and MCRC-V1 ASIC at Stanford\cite{Stueber2025_SPIE}. Error bars indicate the spread of values recorded from multiple readout nodes and datasets of this device. AXIS baseline requirements are shown at 0.5, 1, and 6 keV. The solid line shows the theoretical Fano limit, and the dotted line shows Fano with noise added, as described in the text.}
\label{fig:ccid89_fwhm}
\end{figure}

Low noise is only one ingredient required to achieve excellent spectral response. Charge collection can also be affected by the quality of the entrance window passivation, the level of depletion of the substrate, and charge transfer efficiency. All of these effects are captured in spectral resolution measurements made at several energies across the AXIS energy band using the In-Focus Monochromator (IFM)\cite{Hettrick1990_ifm,Miller2023_AXIS} in the X-ray Detector Lab at MIT. These are shown in Figure \ref{fig:ccid89_fwhm} for two different output speeds, 1 MHz and 2 MHz, which correspond to frame rates of 7 and 14 fps respectively for the 16-channel AXIS detector system. The error bars indicate the spread of values recorded from multiple readout nodes and repeated testing of this device, providing an indication of the range of performance we can expect from similar devices. While these measurements did not incorporate the ASIC, we include a measurement obtained at Stanford at 5.9 keV, using the single-output CCID93 device with the MCRC-V1 ASIC\cite{Stueber2025_SPIE}. All of the 1-MHz (7-fps) spectral FWHM measurements meet AXIS baseline requirements, shown with black bars and arrows. 

The spectral response compares well with theoretical limits. The solid black line in Figure \ref{fig:ccid89_fwhm} shows the fundamental limit imposed by Fano noise---our 5.9 keV measurement with the CCID93 and MCRC ASIC is very close to this. The dotted line shows what to expect if we include 3 e- RMS noise, the AXIS baseline requirement, and if we further assume all events span two pixels. The latter is a reasonable assumption based on previous measurements\cite{Miller2023_AXIS}. The locus of spectral FWHM measurements taken at 1 MHz falls very close to this line, indicating that non-noise contributions to the spectral response are small, if not zero.

\section{Summary}

The High-Speed Camera on AXIS has been designed to exploit the high-throughput, high-spatial-resolution optics and meet the demanding science requirements of this Probe mission concept. Our updated design reflects refinements made during the Phase A study, with several changes made to adapt to the mission's new L2 halo orbit and associated increase in the particle radiation environment. The Front-End Electronics development continues as planned, with a demonstration of full digital waveform sampling of the output video signal expected within the next year.

A first lot of front-illuminated AXIS CCID100 CCDs has completed fabrication, and we expect to begin X-ray performance testing soon. We have updated our parallel test facilities at MIT and Stanford to accommodate these devices, employing a thoughtful modular test bed design to enable efficient, standardized testing of multiple CCDs and 16-channel MCRC ASIC modules. Our recent testing of AXIS prototype CCDs and MCRC ASICs provides confidence that our detector system, and the High-Speed Camera overall, will meet AXIS baseline requirements and enable ground-breaking astrophysics research should AXIS be selected as the next NASA Probe mission.

\acknowledgments 
We gratefully acknowledge support from NASA for the AXIS Probe Phase A study, under contract 80GSFC25CA019, and from NASA SAT grants 80NNSC23K0211 and 80HQTR24TA005. MIT authors acknowledge additional support from the Kavli Institute for Astrophysics and Space Research, and Stanford authors acknowledge support from the Kavli Institute for Particle Astrophysics and Cosmology.

\clearpage


\end{document}